\documentclass[sigconf]{acmart}

\usepackage{booktabs} % For formal tables
\usepackage{amsmath,bm}
\usepackage{enumitem}
\usepackage{footnote}
\usepackage{graphicx}
\usepackage{multirow}
\usepackage{hyperref}
\usepackage{tabularx,ragged2e}
\usepackage{array}
\newcolumntype{L}[1]{>{\raggedright\let\newline\\\arraybackslash\hspace{0pt}}m{#1}}
\newcolumntype{C}[1]{>{\centering\let\newline\\\arraybackslash\hspace{0pt}}m{#1}}
\newcolumntype{R}[1]{>{\raggedleft\let\newline\\\arraybackslash\hspace{0pt}}m{#1}}
\usepackage{threeparttable}
\usepackage{amsmath, amsfonts, amsthm}
\usepackage{xcolor}
\usepackage{booktabs} % For formal tables

\usepackage{booktabs} % For formal tables
\usepackage{amsmath}

\usepackage{algorithm}  
\usepackage{algpseudocode}  
\usepackage{amsmath}  
\usepackage{array}
\usepackage{bm}
\usepackage{balance}
\usepackage[normalem]{ulem}
\usepackage{adjustbox}
\newcolumntype{R}[2]{%
	>{\adjustbox{angle=#1,lap=\width-(#2)}\bgroup}%
	l%
	<{\egroup}%
}
\newcommand*\rot{\multicolumn{1}{R{30}{1.5em}}}

\usepackage{natbib}  % DO NOT CHANGE THIS AND DO NOT ADD ANY OPTIONS TO IT
\usepackage{caption} % DO NOT CHANGE THIS AND DO NOT ADD ANY OPTIONS TO IT
\AtBeginDocument{%
  \providecommand\BibTeX{{%
    \normalfont B\kern-0.5em{\scshape i\kern-0.25em b}\kern-0.8em\TeX}}}

\copyrightyear{2021}
\acmYear{2021}
\setcopyright{rightsretained}
\acmConference[AIES '21]{Proceedings of the 2021 AAAI/ACM Conference on AI, Ethics, and Society}{May 19--21, 2021}{Virtual Event, USA}
\acmBooktitle{Proceedings of the 2021 AAAI/ACM Conference on AI, Ethics, and Society (AIES '21), May 19--21, 2021, Virtual Event, USA}\acmDOI{10.1145/3461702.3462623}
\acmISBN{978-1-4503-8473-5/21/05}
\begin{document}
\fancyhead{} 
\title{Towards Equity and Algorithmic Fairness in Student Grade Prediction}


\author{Weijie Jiang}
\authornote{Both authors contributed equally to this research.}
\affiliation{%
 % \department{Graduate School of Education}
  \institution{University of California, Berkeley}
    \city{Berkeley, CA}
        \country{USA}}
\email{jiangwj@berkeley.edu}

\author{Zachary A. Pardos}
\authornotemark[1]
\affiliation{
%  \department{Graduate School of Education}
  \institution{University of California, Berkeley}
  \city{Berkeley, CA}
      \country{USA}
  }
\email{pardos@berkeley.edu}

\renewcommand{\shortauthors}{Jiang and Pardos}

\begin{abstract}
Equity of educational outcome and fairness of AI with respect to race have been topics of increasing importance in education. In this work, we address both with empirical evaluations of grade prediction in higher education, an important task to improve curriculum design, plan interventions for academic support, and offer course guidance to students. With fairness as the aim, we trial several strategies for both label and instance balancing to attempt to minimize differences in algorithm performance with respect to race. We find that an adversarial learning approach, combined with grade label balancing, achieved by far the fairest results. With equity of educational outcome as the aim, we trial strategies for boosting predictive performance on historically underserved groups and find success in sampling those groups in inverse proportion to their historic outcomes. With AI-infused technology supports increasingly prevalent on campuses, our methodologies fill a need for frameworks to consider performance trade-offs with respect to sensitive student attributes and allow institutions to instrument their AI resources in ways that are attentive to equity and fairness.
\end{abstract}




%\begin{CCSXML}
%<ccs2012>
%  <concept_id>10010520.10010553.10010562</concept_id>
%  <concept_desc>Social and professional topics~User characteristics</concept_desc>
 %</concept>
 %<concept>
 % <concept_id>10010520.10010575.10010755</concept_id>
 % <concept_desc>Computer systems organization~Redundancy</concept_desc>
 % <concept_significance>300</concept_significance>
 %</concept>
 %<concept>
 % <concept_id>10010520.10010553.10010554</concept_id>
  %<concept_desc>Computer systems organization~Robotics</concept_desc>
  %<concept_significance>100</concept_significance>
 %</concept>
 %<concept>
 % <concept_id>10003033.10003083.10003095</concept_id>
 % <concept_desc>Networks~Network reliability</concept_desc>
 % <concept_significance>100</concept_significance>
 %</concept>
%</ccs2012>
%\end{CCSXML}

%\ccsdesc[500]{Social and professional topics~Race and ethnicity}



\maketitle

\section{Introduction}

Equity of outcome, such as degree attainment, is a primary objective of educational institutions. To evaluate how well this goal is being satisfied, administrations, particularly in higher education, often group students by various attributes, such as gender and race, and observe where disparities exist and how long they have perpetuated. An institution may then allocate tutoring and advising types of resources towards groups that exhibit the most disparity in outcome as well as employing curricular redesign and early outreach programs to attempt to address systemic issues that contribute to underachievement. 

AI-infused tutoring \cite{anderson1985intelligent, zhou2020assessing} and advising technology \cite{pardos2019connectionist} is increasingly among the resources an institution has at its disposal to reduce disparities. In higher education, student grade prediction was the first task for which many educational institutions adopted AI to drive school-wide deployment of technological interventions aimed at improving outcomes. Grade prediction was used in early-warning detection systems to flag "at risk" students for faculty and staff to intervene on \cite{harrison2016measuring}, to selectively notify students of available support resources \cite{jayaprakash2014early}, and to directly show live estimates of their chances of passing \cite{arnold2012course}. Nascent campus course information and virtual advising systems \cite{chaturapruek2018data,pardos2019connectionist} are likely candidates to integrate the next generation of grade prediction AI to support personalized recommendation \cite{jiang2019goal}. Secondary education too has seen algorithmic grade prediction become increasingly pervasive and invasive. Final grade predictions of certain students in the United Kingdom, for example, were proposed to take the place of real grades due to the cancellation of exams under COVID-19 \cite{ovchinnikovunethical, opp2020grade}. The proposal was later rescinded after the predicted grades were found to exhibit inaccuracies due to historical biases.

%why race is important
%It has been advocated in the AI, Ethics, and Society community to plan for a \textit{just} AI future in which we endeavor to reduce racial bias \cite{mcilwain2020computerize, addison2019robots}. 

%which resonates with our goal to promote algorithmic fairness and equity in grade prediction from the racial perspective. Especially, the grade prediction algorithm we tend to improve also plays an integral role in a goal-based course recommendation framework \cite{jiang2019goal}, which serves as a pedagogical service to scaffold the learning success of students to degree completion, especially for historically underserved groups that institutions strive to help. Therefore, it is non-trivial to improve the grade prediction model selected by this work towards fairness and equity because it lays a foundation for achieving fairness and equity of future educational outcome, such as course retention and graduation rate.

Fairness and bias in Artificial Intelligence (AI) has attracted substantial attention and developed into a focused research area in the general machine learning community \cite{hardt2016equality, zhang2018mitigating, beutel2019putting}. Endeavoring to reduce racial biases, in particular, has been advocated in the AI, Ethics, and Society community as part of the plan for a \textit{just} AI future \cite{mcilwain2020computerize, addison2019robots}. %\textcolor{red}{Meanwhile, domain-specific educational resources, metrics, processes, and tools are increasingly encouraged given that fairness can be context and application dependent \cite{holstein2019improving}.}
There has been emerging empirical research evaluating fairness in educational contexts with respect to race groups using data analytics \cite{yu2020towards, gardner2019evaluating}; however, no work has yet focused on improving educational equity and fairness from an AI perspective. In this work, we present methodologies for evaluating fairness of grade prediction with respect to race, then design for equity \cite{gutierrez2016social} with a novel boosting of underserved groups based on historic graduation outcomes. Our empirical results are based on institution-wide course grades and demographics from a large public university. We propose strategies during the data processing stage, the model training stage, and the inference (prediction) stage of the grade prediction model to improve group fairness while maintaining overall accuracy. Experiment results demonstrate that: (1) adversarial learning produces the highest fairness scores while leading to minimal overall reduction in prediction performance, (2) our proposed equity-based strategy is largely effective as most of the underserved groups exhibit higher average improvements than other groups in all three evaluation metrics, and (3) the most performant model strategies vary for different race groups.%, raising the question of whether a race group-specific or single universal model should be adopted.

\section{Related Work}

\subsection{Fairness in Machine Learning and Education}
\label{section_metrics}
The dramatic progress of AI has led to machine learning algorithm adoption in many high-stake applications, including employment, criminal justice, personalized medicine, and education \cite{gajane2018formalizing}. Nevertheless, fairness in machine learning remains a problem in that machine learning algorithms risk amplifying social inequities by over-associating sensitive attributes (e.g., race and gender) with prediction labels, which may lead to discriminatory behaviors against certain subgroups \cite{ali2019loss, wang2019balanced, alvero2020ai}, such as women in the STEM workforce \cite{kiritchenko2018examining}. 

Many metrics have been proposed to measure group fairness. %One of the simplest conceptions of fairness is the notion of \textit{demographic parity}. 
\textit{Demographic parity} requires that, for all groups of a sensitive attribute (e.g. race), the overall probability of a positive prediction of a given outcome should be the same - the sensitive attribute should be independent of the prediction \cite{calders2009building}, i.e. $P(Y'=k|A=i)=P(Y'=k|A=j)$, where the model prediction is $Y'$ and the sensitive attribute is denoted by $A$. However, the usefulness of demographic parity can be limited if the base rates of the two groups differ, i.e. if $P(Y=k|A=i)=P(Y=k|A=j)$, where $Y$ represents the ground truth. Two alternative criteria were developed by conditioning the metric on $Y$, yielding \textit{equalized odds} and \textit{equal opportunity} \cite{hardt2016equality}. Equal odds requires equal true positive rate and false positive rate between the groups, formally, $P(Y' =1|A=i,Y=y)=P(Y'=1|A=j,Y=y), \forall y \in \{0, 1\}$.
Equal opportunity requires only one of these equalities and is intended to match errors in the “advantaged” outcome, such as "admission to college", across groups, formally, $P(Y' =1|A=i,Y=1)=P(Y'=1|A=j,Y=1)$.
%To get rid of the impact of threshold on calculating true positive rate and false positive rate, \citet{gardner2019evaluating} proposed Absolute-Between-ROC Area (ABROCA) for quantifying how a predictive model’s performance varies across different student subgroups by comparing the group’s ROC curve to the ROC curve of a baseline group.  %In a study analyzing MOOC dropout rate, they show a significant difference in privilege given to males versus  females in machine learning models across a variety of feature sets and across a classification techniques.

In education, considerations of fairness are deeply rooted and focused on concerns of bias and discrimination \cite{kizilcec2020algorithmic}. With the increasing use of data and machine learning models in educational technologies to provide support and analytic insights to students, instructors, and administrators, problems arise in terms of its impact on fairness in an education system. 
For example, on-time college graduation prediction from application data can treat certain subgroups of students unfairly and cause less accurate predictions for them \cite{hutt2019evaluating}. A machine learning based predictor may underestimate underrepresented demographic groups when predicting college student success \cite{yu2020towards}, and many grade prediction approaches cannot achieve good accuracy in predicting underachieving students \cite{polyzou2019feature}.
 The fairness problem in education may cause adverse impacts on individuals and society by not only constraining a student’s opportunity, but also exacerbating historic social inequities.
However, formalized research on improving algorithmic fairness in educational technologies has been limited.
It is therefore essential to take into account fairness (i.e., equity of opportunity) in decision support algorithms used in education, so as not to suppress hope of students by closing off paths due to algorithmic bias. 

\subsection{Fairness Problem Categorization} 
\label{problem}
Fairness problems can be generally categorized into two classes from computational perspective: prediction outcome discrimination due to high feature-class correlation and prediction quality disparity due to imbalanced data \cite{du2020fairness}. 

%Prediction Outcome Discrimination due to high feature-class correlation: Discrimination refers to the phenomenon that DNN models produce unfavourable treatment of people due to the membership of certain demographic groups. 
Because of the intrinsic noise or additional signals of certain high feature-class correlation that commonly exist in data, machine learning models would naturally replicate the biases in the skewed data and eventually result in algorithmic bias.
Even though a machine learning model that excludes sensitive attributes from model input attempts to achieve fairness through unawareness, it may still induce prediction discrimination because a learned model can inadvertently reconstruct sensitive attributes from a number of seemingly unrelated features \cite{kizilcec2020algorithmic}. For instance, ZIP code and surname could indicate race. The model prediction might highly depend on the class memberships, and eventually show
discrimination to certain demographic groups \cite{kizilcec2020algorithmic}.

Given that the typical objective of training a machine learning model is to minimize the overall error but usually the training data may be less informative for certain parts of the population, if the model cannot simultaneously fit all populations optimally, it will fit the majority group. Although this may maximize the overall prediction accuracy, it might come at the expense of underrepresented populations and lead to poor performance for those groups.
For example,    
\citet{yu2020towards} showed that the imbalanced student subpopulations could be the main source of inequalities and unfairness in predicting academic success of college students. \citet{doroudi2019fairer} demonstrated that knowledge tracing algorithms could also be inequitable, favoring fast learners over slow learners, when using student models that are fit to aggregate populations of students.

\subsection{Mitigation of Algorithmic Bias}
\label{mitigate}

Strategies to mitigate algorithmic bias can be designed and implemented in the three stages of a typical machine learning pipeline: dataset construction, model training, and inference. 

It is a straightforward solution during the dataset construction stage to remove fairness sensitive features from training data. %For instance, surname and ZIP code can be deleted to reduce the discrimination of machine learning models towards certain race. 
However, prediction outcome discrimination may still be perpetuated because of other feature-class correlation. Directly removing features might also lead to poor model performance \cite{pedreshi2008discrimination}. For example, it was shown that disregarding the race feature of students harms both overall accuracy and demographic parity of an algorithmic admissions system that predicts college success \cite{kleinberg2018algorithmic}. For a system that predicts learning outcomes of university students using data from a learning management system, predictions become more accurate if the feature set includes student demographic information \cite{yu2020towards}. 
Further techniques to ensure fairness in the data construction stage include assigning different weights to training samples \cite{kamiran2012data} and re-weighting each label for the loss function, which are targeted for imbalanced data in terms of group and predicted class, respectively. %Feature swapping was also used to create a dataset which is identical to the original one but biased towards certain values of the feature in order to be merged with the original data to achieve balance in terms of that feature.
However, even when training data is balanced, machine learning models may still capture information like gender and race in intermediate representations \cite{wang2019balanced}. %Thus it is essential to design further techniques in the other two stages to reduce discrimination in machine learning models.
%Since prediction discrimination is partially caused by difference of label distribution conditioning on sensitive features in the training data, feature swapping can also be used to create a dataset which is identical to the original one but biased towards certain value of the feature. The union of the original and feature-swapped dataset would be balanced in terms of the that feature. 
\begin{figure*}[h]
	\centering
	\includegraphics[width=0.85\linewidth]{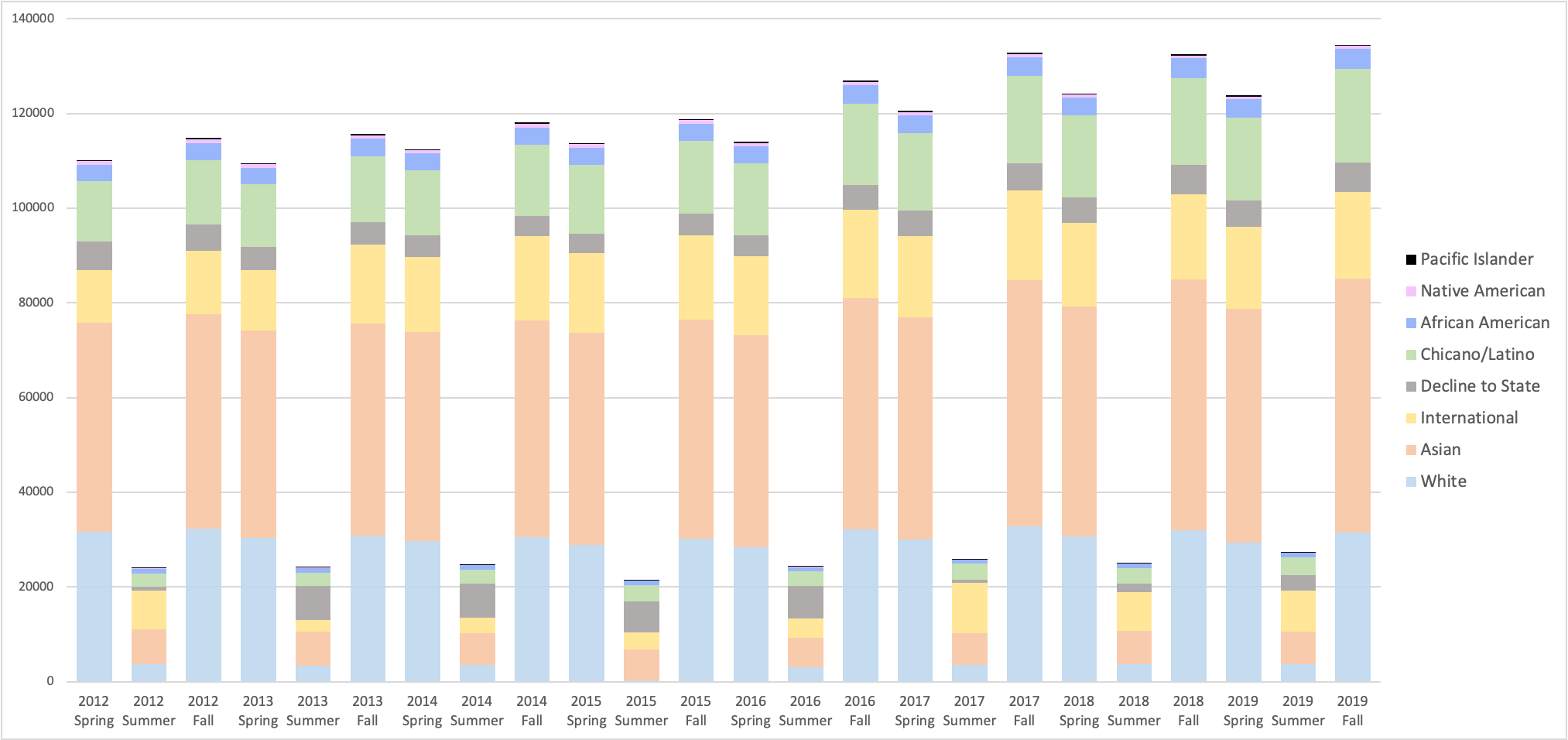}
	\caption{Distribution of enrollments by race across semesters}
	\label{race_enrollments}
\end{figure*}

Adversarial learning has been leveraged to reduce modeling bias during training, removing information about sensitive attributes from intermediate representation of model input in predictive models \cite{beutel2017data, zhang2018mitigating, madras2018learning,
wadsworth2018achieving, wu2020fairness}. In adversarial learning, a predictor and an adversarial classifier are learned simultaneously. The goal of the predictor is to ensure the representations of model input are maximally informative for the major prediction task, while the adversarial classifier is designed to minimize the predictor’s ability to predict the sensitive attribute \cite{du2020fairness}. Thus, adversarial learning has the potential to learn bias-free representations of model input by removing the bias information about sensitive user attributes.
The mitigation of modeling bias could also be implemented at the inference stage. The key idea is to suppress the parts of the model that have captured sensitive attributes so as to turn off the correlation between those attributes and model predictions \cite{du2020fairness}.

\section{Datasets}
\label{datasets}
\subsection{Student Enrollment Data}
We used a novel dataset from UC Berkeley, a large public liberal arts university in the US, which contained anonymized student course enrollments from Spring 2012 through Fall 2019. The dataset consisted of per-semester course enrollment information for 82,309 undergraduates with a total of 1.97 million enrollments. A course enrollment meant that the student was still enrolled in the course at the conclusion of the semester. The median courses enrolled in per semester was four. Student course scores consisted mostly of letter grades (i.e., A, B, C, D, F) with some courses allowing students to elect to be graded based on a PASS/No-PASS score, a passing grade being equivalent to a C- or higher. 
There were 10,430 unique courses, including 9,714 unique primarily lecture courses from 197 subjects in 124 different departments hosted in 17 different divisions of 6 colleges. In all analyses in this paper, we only considered lecture courses with at least 20 enrollments total over the 8 year period. The raw data were provided in CSV format by the University's Enterprise Data and Analytics unit.

\subsection{Student Demographic Data}
In addition to student enrollment data, the dataset also contained  demographic information of students, including their gender, race, entry status, and parental income when admitted. Racial subcategories listed were: White, Asian, International, Chicano/Latino, African American, Native American/Alaskan Native, Pacific Islander, and Decline to State. Chicano/Latino, African American, Native American/Alaskan Native, and Pacific Islander students are currently underrepresented at the University. %UC Berkeley\footnote{\url{https://diversity.berkeley.edu/reports-data/diversity-data-dashboard}}.

\subsection{Descriptive Analyses by Race Group} 
Enrollments for each semester, broken out by race, is shown in Figure \ref{race_enrollments}. Enrollments by Asian, White, and International students rank in the top three, accounting for 77.42\% of all enrollments, with the four underrepresented groups accounting for 17.03\% of the enrollments, and 5.55\% from students declining to state their race.

\begin{figure}[h]
	\centering
	\includegraphics[width=0.8\linewidth]{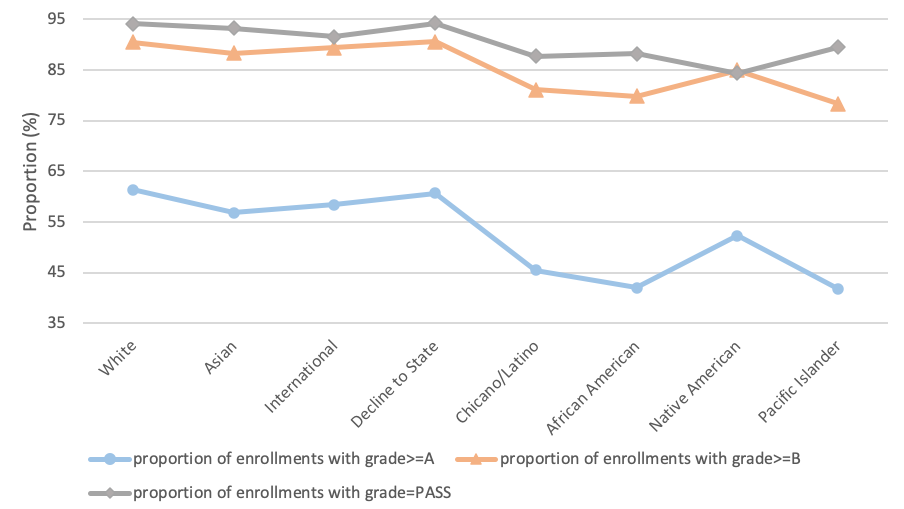}
	\caption{Grade distribution by race}
	\label{grade_distribution}
\end{figure}

Figure \ref{grade_distribution}
depicts the proportion of course grades in the A category (including A-, A, and A+), not lower than B, and the proportion of PASS grades among all the non-letter grades (i.e., including only PASS and No-PASS grades). Generally, the proportion of enrollments graded with PASS is much higher than those with A for all groups. Among all the enrollments with non-letter grades by White, Asian, and International students, $90\%\sim95\%$ of them were PASS, compared with $85\%\sim90\%$ of those were PASS by Chicano/Latino, African American, Native American/Alaska Native and Pacific Islanders. Additionally, $55\%\sim65\%$ of  enrollments with letter grades by White, Asian, and International students were in the A category, compared with $35\%\sim55\%$ of the enrollments by Chicano/Latino, African American, Native American/Alaska Native, and Pacific Islanders with the same grade type. A similar pattern exists for proportions of the not lower than B category.

A growing literature points to opportunity gaps at a systemic level as leading to these observed achievement gaps among student groups, many from underresourced communities \cite{carter2013closing}. While it can be difficult to explicate these disparities, acknowledging the presence of racial inequity is a necessary first step towards better serving the historically underserved \cite{carter2017you}.

%Based on the descriptive analysis above, Chicano/Latino, Afican American, Native American/Alaska Native, and Pacific Islanders are relatively underrepresented in terms of both population and learning success, who therefore deserves more help in the learning process. 

\section{Course Grade Prediction with LSTM}
Long Short-Term Memory (LSTM), a popular variant of RNNs, has been used to good effect as a dynamic course grade prediction model \cite{jiang2019goal, jiang2019time}. To prepare our dataset for training this model, enrollment grade sequences, $\bm{g}_t$, and course sequences, $\bm{c}_t$, of a student are converted to fixed length input vectors,
\begin{equation*}
\bm{g}_t = (\bm{g}_t^1, \bm{g}_t^2, ..., \bm{g}_t^n) 
\end{equation*}
\begin{equation*}
\bm{g}_t^i = (s_{ti}^1, s_{ti}^2, ..., s_{ti}^m, s_{ti}^{Pass}, s_{ti}^{No-Pass})
\end{equation*}
\begin{equation*}
\bm{c}_t = (c_t^1, c_t^2, ..., c_t^n)
\end{equation*}
where $n$ denotes the number of courses, $m$ denotes the number of letter grades that students can receive for a course, and $t$ is the time tag for semester. Therefore, $\bm{g}_t^i \in \{0, 1\}^{m+2}$, $\bm{g}_t \in \{0, 1\}^{(m+2)*n}$, and $\bm{c}_t \in \{0, 1\}^{n}$. \citet{jiang2019goal} showed that using previous semester's course grades and current semester's enrollments as input to the hidden layer of LSTM always achieved better grade prediction performance than only using the previous semester's grades. In order to separate the loss calculated from the letter grades and PASS/NO-PASS grades and mask the semesters that students did not enroll in, a two-level masked cross-entropy loss function was specified as:

\begin{equation}
\begin{aligned}
\small
L_{masked}&= MaskedCrossEntropy(\hat{\bm{g}}_{t+1}, \bm{g}_{t+1})\\ 
&=- \sum_t \sum_{i, \hat{\bm{g}}_{t+1}^i\neq \bm{0}} (\hat{\bm{g}^{i1}}_{t+1}^T \textrm{log} \bm{g}_{t+1}^{i1}+\hat{\bm{g}^{i2}}_{t+1}^T \textrm{log} \bm{g}_{t+1}^{i2})
\label{loss0}
\end{aligned}
\end{equation}
where $\bm{g}_t^{i1} = (s_{ti}^1, s_{ti}^2, ...,s_{ti}^m)$, $\bm{g}_t^{i2} = (s_{ti}^{Pass}, s_{ti}^{No-Pass})$, and $\hat{\bm{g}^{i1}}_{t}$ and $\hat{\bm{g}^{i2}}_{t}$ denote the ground truth of grade (i.e., the labels for training the grade prediction model). 

%Therefore, the LSTM output at each time slice (semester) is calculated by $\bm{h}_{t-1}, \bm{g}'_t = \textit{LSTM}([\bm{g}_{1}, \bm{c}_{2}],...,[\bm{g}_{t-1}, \bm{c}_{t}],\bm{h}_0)$

\section{Strategies to Mitigate Bias in Grade Prediction}
\label{strategy}
We will employ and adapt strategies for mitigating algorithmic bias, as referred to in related work, to the grade prediction task. These strategies can be utilized in three stages of the LSTM prediction pipeline: data construction, model training, and inference. We summarize all the strategies which will be described in this section in Table \ref{method_summary}.
\begin{table}[h]
	%\centering
	\small
	\caption{Summary of strategies which will be used to attempt to mitigate bias in the LSTM grade prediction model}
	\label{method_summary}
	\begin{tabular}{L{3cm}|L{2cm}|L{2cm}l}
		\toprule
		\textbf{Strategy} &\textbf{Name}  &
		\textbf{Stage} \\
		\midrule[0.08em]
		fairness through unawareness &default (loss)& -\\ \hline
		weight loss by grade label & grade label weighted loss & data construction\\
		\hline
		weight loss by sample &alone, grad-rate (wgh), equal (wgh) & data construction\\
		\hline
		sensitive feature added to input & race (feature) & data construction\\ \hline
		multiple features added to input &multi & data construction\\ \hline
		adversarial learning & adversarial  & model training\\ \hline
		remove features for prediction &infer-rmv & inference (prediction)\\
		
		\bottomrule

	\end{tabular}
\end{table}

\begin{figure}[h]
	\centering
	\includegraphics[width=0.8\linewidth]{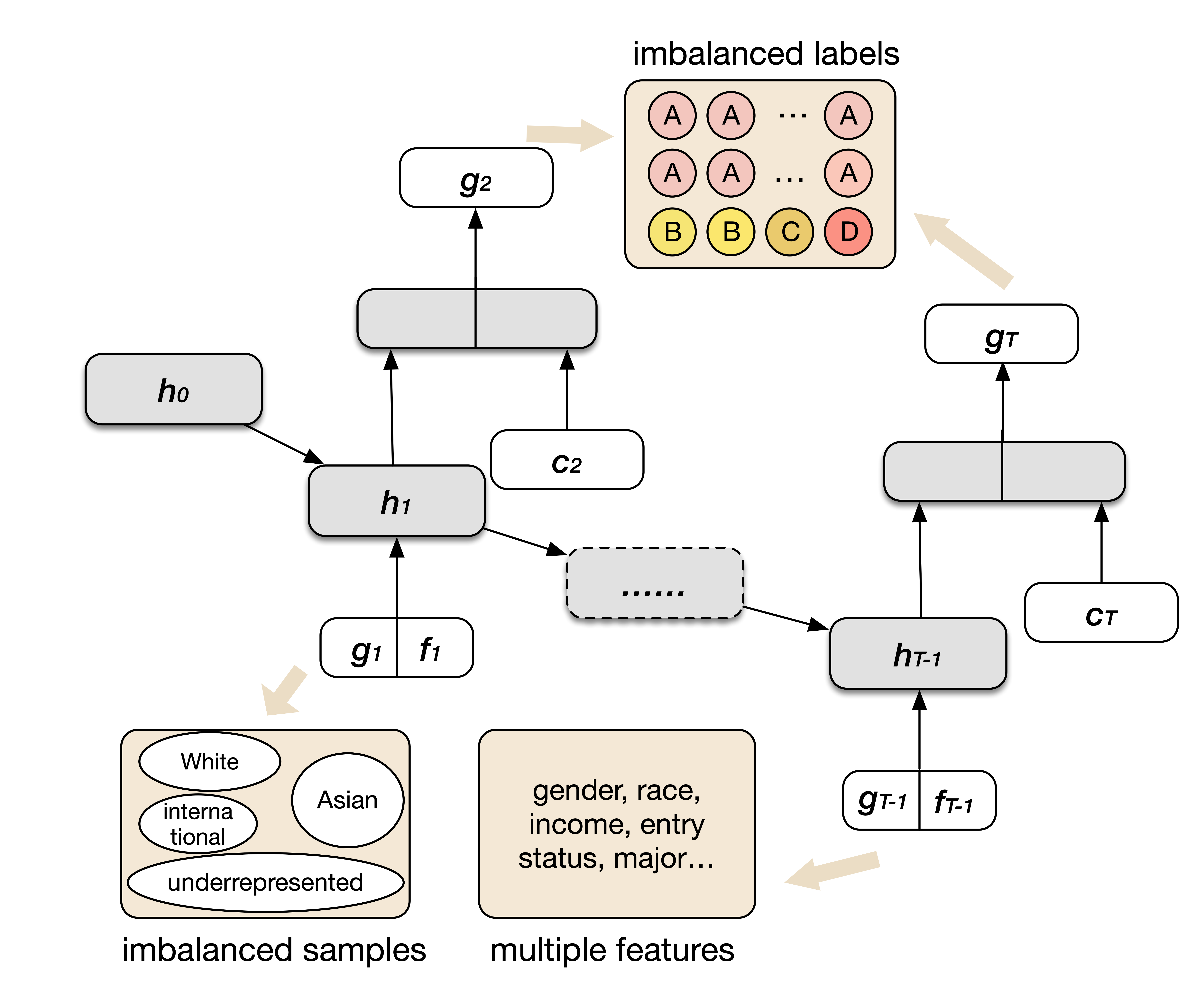}
	\caption{Pre-processing strategies to improve fairness}
	\label{RNN}
\end{figure}

\subsection{Data Construction Strategies}
\label{pre-processing_strategy}
Figure \ref{RNN} illustrates the LSTM grade prediction framework and three factors in the data construction stage that may introduce bias into the prediction model.

First, training samples can be very imbalanced with respect to sensitive student attributes, which may lead to the prediction quality disparity problem as is mentioned in the "Related Work" section. To deal with the issue of imbalanced data samples and aim for fairness in the data construction stage with respect to race, we can balance the influence of training samples in the loss function by assigning weights to counteract racial underrepresentation \cite{kamiran2012data}. The adjusted loss function of the LSTM grade prediction is expressed as:
\begin{equation}
\small
L_{wgs} = -\sum_t \lambda(r(\hat{\bm{g}}_{t+1}))MaskedCrossEntropy(\hat{\bm{g}}_{t+1}, \bm{g}_{t+1})
\label{wgh_sample_eq}
\end{equation}
where $r(\hat{\bm{g}}_{t+1})$ denotes the race of the student sample that has the grade label $\hat{\bm{g}}_{t+1}$, and $\lambda(r(\hat{\bm{g}}_{t+1}))$ assigns the student sample with a specific weight associated with their race. Normally, the form of $\lambda(*)$ varies, but the weights associated with majority groups should be set smaller than those of minority groups in the data, so as to give the model greater chances to learn from the less represented groups. Scenarios may occur When the weight of a certain race group is set much larger than that of other groups, then the model will mainly learn from that group and ignore samples from other groups. 

%It is worth mentioning that with historic disparities in achievement perpetuated among students for a long time, 
Even if the number of instances are the same across groups, an institution may still like to utilize a "weight loss by sample" strategy to mitigate historic equity gaps, such as differences in graduation-rate with respect to racial, gender, or socioeconomic groupings. We introduce this equity oriented weighting, in which group weights are set in negative correlation to their historic outcomes. This can also be applied when instances are not balanced, by overrepresenting groups with lower historic outcomes, which may also be underrepresented, instead of bringing them to parity. This is an example of equity by design \cite{gutierrez2016social}, where the design and efficacy of an intervention is centered around non-dominant groups. %In the context of an educational technology intervention using grade prediction, it also means designing the technology and tuning the algorithm to prioritize predictive accuracy on underserved groups.

%Imbalance with respect to groupings can exist outside of instances.
A second factor that may introduce bias is imbalanced label distribution across groups. Student grade label distributions in colleges and universities have been reported to exhibit inflation, narrowing, and unevenness \cite{weaver2007more, matos2010student, polyzou2019feature,arthurs2019grades},  also reflected in our dataset as illustrated in Figure \ref{grade_distribution} where roughly half of grades are in the $A$ category and more than $80\%$ are $B$ or better. When differences between group grade distributions exist and there are significant group size differences, a model is likely to be biased towards the distribution of the largest groups, worsening the grade prediction fairness problem. %Regardless of the reasons behind the uneven distributions of grades, it may worsen the prediction fairness problem due to the imbalanced labels among underserved groups.
Similar to instance balancing, we can balance labels by giving different weights to training samples based on their grade labels, with a resulting adjusted loss function of the LSTM grade prediction defined as:

%\begin{equation}
%\small
%L_{wbl} = \sum_t %\sigma(\hat{\bm{g}}_{t+1})MaskedCrossEntropy(\hat{\bm{g}}_{t+1}, \bm{g}_{t+1})
%\label{wgh_label}
%\end{equation}

\begin{equation}
\small
  L_{wbl} =  - \sum_t  \sum_{i, \hat{\bm{g}}_{t+1}^i\neq \bm{0}} \sigma(\hat{\bm{g}^{i}}_{t+1}) (\hat{\bm{g}^{i1}}_{t+1}^T \textrm{log} \bm{g}_{t+1}^{i1}+\hat{\bm{g}^{i2}}_{t+1}^T \textrm{log} \bm{g}_{t+1}^{i2})
  \label{wgh_label}
\end{equation}
where $\sigma(\hat{\bm{g}}^i_{t+1})$ assigns each enrolled course of a student sample with a specific weight according to its grade label $\hat{\bm{g}}^i_{t+1}$. In the case of using both a race group representation-based instance balancing with label-based balancing, the two weighting schemes are combined as defined by:

\begin{equation}
\tiny
  L_{wbsl} = - \sum_t \lambda(r(\hat{\bm{g}}_{t+1}))\sum_{i, \hat{\bm{g}}_{t+1}^i\neq \bm{0}} \sigma(\hat{\bm{g}^{i}}_{t+1})  (\hat{\bm{g}^{i1}}_{t+1}^T \textrm{log} \bm{g}_{t+1}^{i1}
  +\hat{\bm{g}^{i2}}_{t+1}^T \textrm{log} \bm{g}_{t+1}^{i2})
  \label{wgsl_label}
\end{equation}

Third, the "fairness through unawareness" strategy has been demonstrated to be ineffective because it falls short of being blind to sensitive attributes as they can be inadvertently reconstructed from a number of seemingly unrelated features \cite{dwork2012fairness, kizilcec2020algorithmic, kleinberg2018algorithmic, yu2020towards}. Instead, sensitive student attributes, such as gender, race, and family income, should be acknowledged and modeling strategies employed to mitigate any bias introduced by them. A first step is to present sensitive attributes, $\bm{f}_t$,  along with grade information to the model (Figure \ref{RNN}). The feature embeddings learned by the LSTM might take away sensitive attribute-related information from the grade embeddings and enable them to be less biased from those attributes.

\subsection{Model Training Strategy with Adversarial Learning}
\label{advsarial_method}
Adversarial learning is a technique that has been used to attempt to learn bias-free deep representations from biased data \cite{zhang2018mitigating, beutel2017data, madras2018learning, wu2020fairness}. Its mission is to enforce the deep representations to be maximally informative for predicting the labels of the main task while minimally discriminative for predicting sensitive attributes \cite{du2020fairness}. We start with the LSTM grade prediction model that outputs a probability distribution of grades for each course student took in a semester. %The goal is to ensure the model prediction to satisfiy equalized odds or equal opportunity for a sensitive feature $f$, such as race. %Even if the sensitive feature is not an input feature to our LSTM, it may be correlated with the class/label, which might cause the model to be biased. 
\begin{figure}[h]
	\centering
	\includegraphics[width=0.85\linewidth]{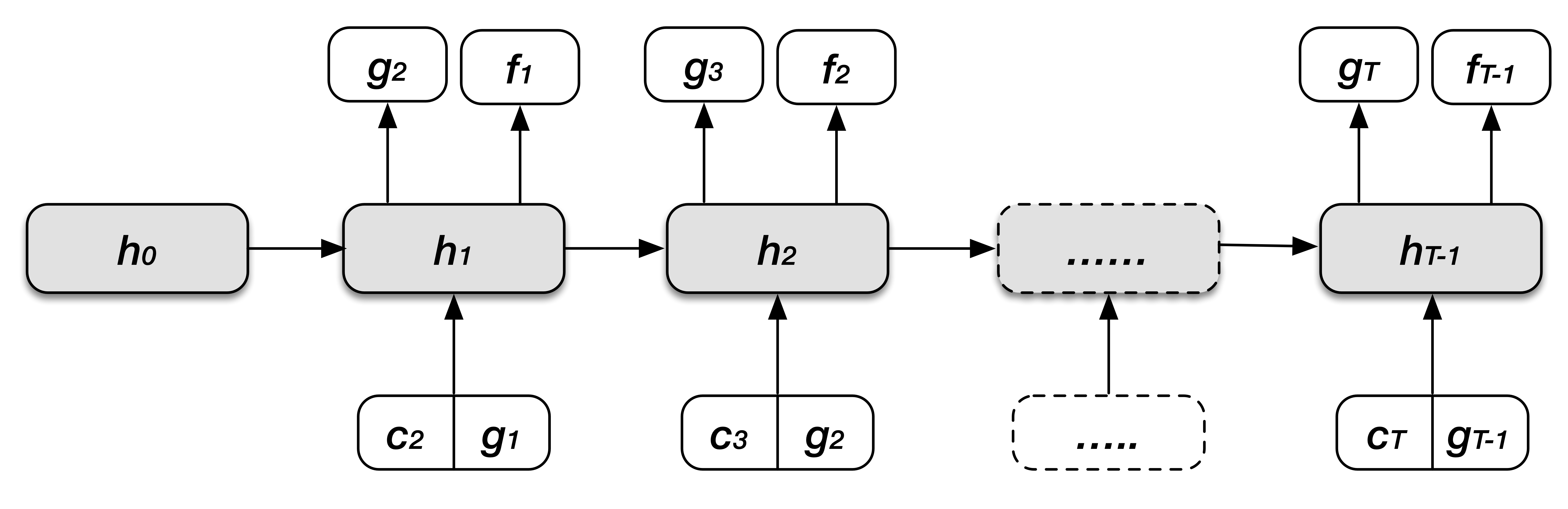}
	\caption{LSTM grade prediction framework with adversarial learning }
	\label{RNN2}
\end{figure}
The goal in this scenario is for the LSTM to be accurate at the task of predicting student grades while maintaining maximum uncertainty with respect to the race of the student. A straightforward approach is to apply an attribute discriminator to the hidden states learned by the LSTM to infer race and penalize the model according to the negative gradients from the adversarial loss that indicates the informativeness of hidden states for race prediction. 
We add another output layer on top of the hidden states in the LSTM model to predict the sensitive attribute of race $\bm{f}_t$ at time slice $t$, as illustrated in Figure \ref{RNN2}. The adversarial loss function for predicting the categorical sensitive attribute is cross-entropy, which is formulated as:
\begin{equation}
Loss_F = - \sum_t \hat{\bm{f}}_{t}^T \textrm{log} \bm{f}_{t}
\label{loss}
\end{equation}
If we subtract $loss_F$ from the original masked cross entropy loss of the LSTM grade prediction model, which is formulated in (\ref{loss0}), the model will be encouraged to maximize $loss_F$, which will prevent the learned course grade embedding and hidden states of the model from being able to predict race accurately. Meanwhile, the model still has to maintain the ability to predict grades, therefore, the weight of the two losses needs to be tuned so as not to harm the grade prediction performance unnecessarily. The final loss function is formulated as:
\begin{equation}
\small
L = - \sum_t \sum_{i, \hat{\bm{g}}_{t+1}^i\neq \bm{0}} (\hat{\bm{g}^{i1}}_{t+1}^T \textrm{log} \bm{g}_{t+1}^{i1}+\hat{\bm{g}^{i2}}_{t+1}^T \textrm{log} \bm{g}_{t+1}^{i2}) + \alpha\sum_t \hat{\bm{f}}_{t}^T \textrm{log} \bm{f}_{t}
\label{loss3}
\end{equation}
where $\alpha$ is a coefficient that controls the importance of the adversarial loss function.

\subsection{Inference Strategy}

An additional strategy we trial towards achieving fairer grade prediction is to use sensitive attributes in training, but not in the inference (i.e., prediction) stage. For the LSTM model that takes in sensitive attributes concatenated with grades as model input, illustrated in Figure \ref{RNN}, we hypothesize that the feature embeddings learned by the LSTM might take away some sensitive attribute-related information from the grade embeddings and enable the grade embeddings to be less biased based on sensitive attributes. Therefore, in the inference stage of grade prediction, we can attempt to remove feature information from the input by only giving the historical grades to the model input.

\begin{figure*}[h]
	\begin{minipage}{0.33\textwidth}
		\centering
		\includegraphics[width=1\linewidth]{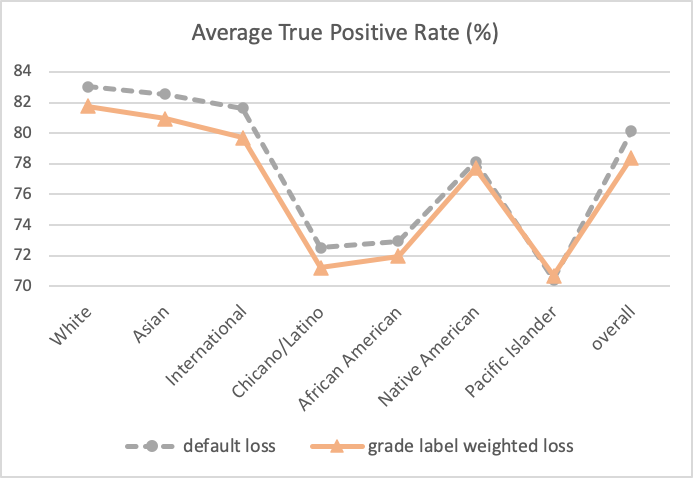}
		%\caption{Interpolation for Data 1}\label{Fig:Data1}
	\end{minipage}\hfill
	\begin{minipage}{0.33\textwidth}
		\centering
		\includegraphics[width=1\linewidth]{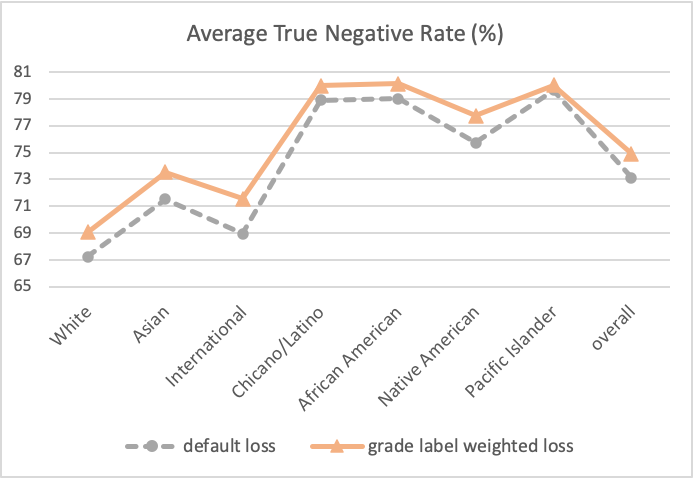}
		%\caption{Interpolation for Data 2}\label{Fig:Data2}
	\end{minipage}\hfill
	\begin{minipage}{0.33\textwidth}
		\centering
		\includegraphics[width=1\linewidth]{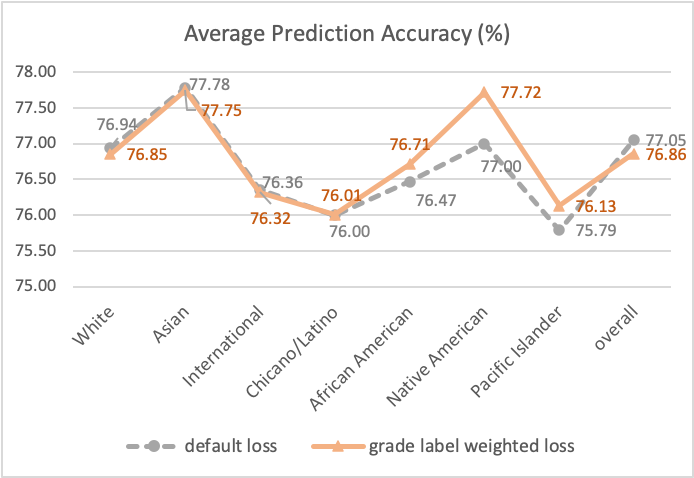}
		%\caption{Interpolation for Data 2}\label{Fig:Data2}
	\end{minipage}
	\caption{Results of comparison between models with unweighted loss and models with weighted loss by grade label}
	\label{avg_compare}
\end{figure*}

\section{Experiment Results Analysis}
\label{post_strategy}

In this section, we evaluate the proposed strategies in terms of model fairness and equity, where the metrics of accuracy, true positive rate, and true negative rate are selected to be reported according to \textit{equity of odds} and \textit{equity of opportunity} that are mentioned in related work. Accuracy measures the overall predictive power of the model. If we set a cutoff for letter grades to divide them into two groups, such as "not lower than $A(B)$" and "lower than $A(B)$", then true positive rate (TPR) reflects the probability of predicting well-performing students, which can be a measure of equal opportunity among groups when an intervention uses a predicted high grade to open up opportunities for students  \cite{hardt2016equality}. %  Ideally, we hope that same proportion of well-performing students can be predicted correctly from each race group of students so that they can have the same opportunity to be honored.
A high false negative rate (FNR) can lead to reduced opportunities in the form of a hypothetical grade-based intervention “underplacing” or unjustly precluding students from opportunities. True negative rate (TNR), on the other hand, captures the possibility that students who need help for their studies can be accurately detected. While in \citet{hardt2016equality}, they select equal TPR to represent equity of opportunity, in our context we also consider TNR as it represents equity of opportunity to be helped. These metrics can shed light on potential consequences of using grade prediction in different applications. In this work, we set the grade category $A$ as the cutoff for binary grade prediction due to the grade distribution that enrollments with grades in the $A$ category take up around $56.12\%$ of the overall enrollments in our data. Students who cannot receive an $A$ can be deemed as scoring behind half of the students on average. We used the datasets introduced in the "Datasets" section for experiments, where data from 2012 Spring to 2018 Summer are used for training, 2018 Fall for validation, and 2019 Fall as the test set. The size of the training, validation, and test data are in the proportion of 13:1:1.

\subsection{Debias the Imbalanced Grade Labels}
\label{debias-label}

Given the uneven distribution of grades in the whole data population with $56.12\%$ not lower than $A$, as well as the disparities in grade distributions among groups, we evaluate how weighting the loss function by grade label (i.e., balancing by grade label) mitigates the prediction quality disparity problem due to the imbalanced labels. Specifically, we trained the model by minibatch, calculating $\sigma(\hat{\bm{g}}^i_{t+1})$ in equation (\ref{wgh_label}) based on the proportion of each type of label (i.e., grade type) in each minibatch, $\sigma(\hat{\bm{g}}^i_{t+1}) = \frac{1/p(\hat{\bm{g}}^i_{t+1})}{\sum_t 1/p(\hat{\bm{g}}^i_{t+1})}$, where function $p$ calculates the proportion of the grade type that a student received for the $i$-th course, i.e., $\hat{\bm{g}}^i_{t+1}$, in each minibatch.

Figure \ref{avg_compare} shows a comparison of average results in terms of the three metrics between models with unweighted loss and models with weighted loss by grade label. All the values are averaged based on the results of all the strategies listed in Table \ref{method_summary}. Overall, models with unweighted loss tended to achieve higher TPR than TNR on average ($80.12\%$ v.s. $73.13\%$), which is largely because the model has fit the larger proportion of samples with grade label $A$ better than the other group of samples with grade lower than $A$. %This reflects inequity between well-performing students and students for whom the institution may mobilize resources to support.%, because more students in need will be missed out by the model due to smaller TNR (larger false positive rate).
After adopting the weighted loss by grade label, the gap between overall TPR and TNR became narrower ($78.39\%$ v.s. $74.91\%$).  
When splitting the whole student population by race, it is apparent that the model achieved higher TPR but lower TNR for White, Asian, and International students than Chicano/Latino, African American, Native American, and Pacific Islander students, likely due to the larger proportion of students with $A$ in the first race groups. Models with weighted loss function by grade label also decreased the TPR and increased TNR for all race groups, with changes more salient for the first three groups, meaning the unfairness problem between well-performing students and underachieving students within each race group has been mitigated to some degree. The average prediction accuracy results show that weighting the loss function by grade label boosted accuracy for Chicano/Latino, African American, Native American, and Pacific Islander students without sacrificing much accuracy for White, Asian, and International students \footnote{The overall decrease of accuracy might result from the much larger population of the first three race groups than the other four.}. Therefore, we consider weighting the loss by grade label in training as an efficient strategy to debias the imbalanced grade labels, and apply it to all models introduced in subsequent analyses.  

\subsection{Debias the Imbalanced Race Groups}
\label{debias_imb_races}

Sample re-weighting based on student race can be a solution to deal with the problem of imbalanced race groups, which aims at giving underrepresented groups larger representation in training by weighting the loss function to draw more emphasis from the model. Without sample re-weighting, each sample is given the same weight in the loss function, but the majority groups will attract more attention from the model training process because they have more samples than lesser represented groups. We use "default" to denote this strategy because it used the default loss function, equation (\ref{loss0}), which is the same as the "fairness through unawareness" strategy we mentioned in section "Strategies to Mitigate Bias in Grade Prediction with LSTM". 

In order to assign lesser represented groups with larger weights and the vice versa, we define the weighting function $\lambda$ in equation (\ref{wgh_sample_eq}) as $\lambda(\bm{r})=1/\bm{r}$, where $\bm{r}$ is a proportion vector of enrollments by each race group in the data. Therefore, after re-weighting, each race will share equal weight in the loss function on average. This strategy is denoted by "equal".

If we consider a curricular recommender system in which the grade prediction model affects the quality and success of a student's curricular path, such as on-time graduation in higher education, an institution may elect for the efficacy of the recommender system to be boosted for historically underserved groups even at the expense of the efficacy on groups with historically high graduation rates. 
\begin{figure*}[h]
	\begin{minipage}{0.33\textwidth}
		\centering
		\includegraphics[width=1\linewidth]{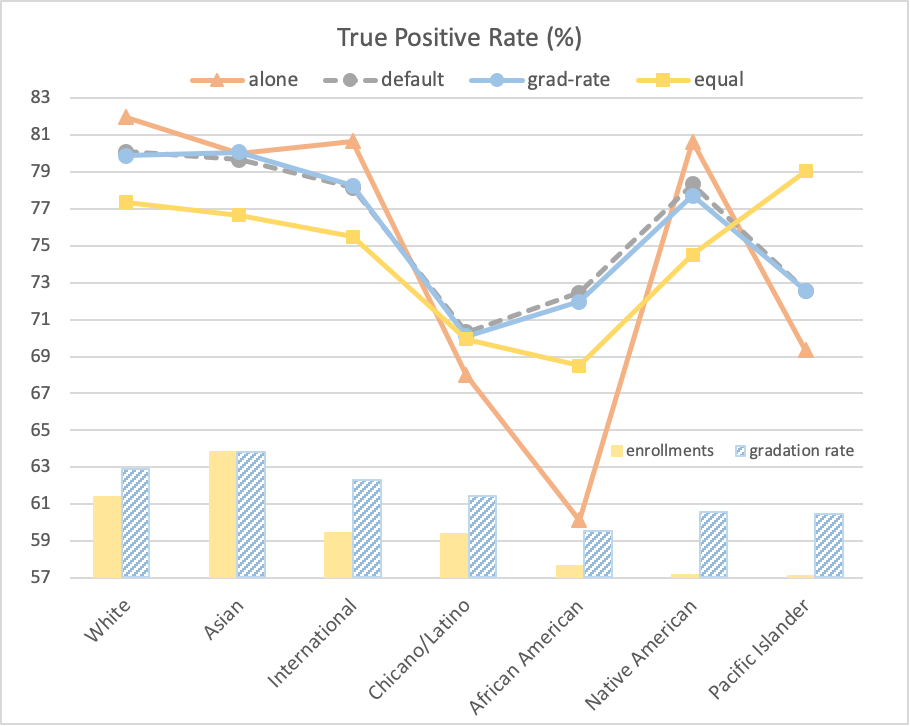}
		%\caption{Interpolation for Data 1}\label{Fig:Data1}
	\end{minipage}\hfill
	\begin{minipage}{0.33\textwidth}
		\centering
		\includegraphics[width=1\linewidth]{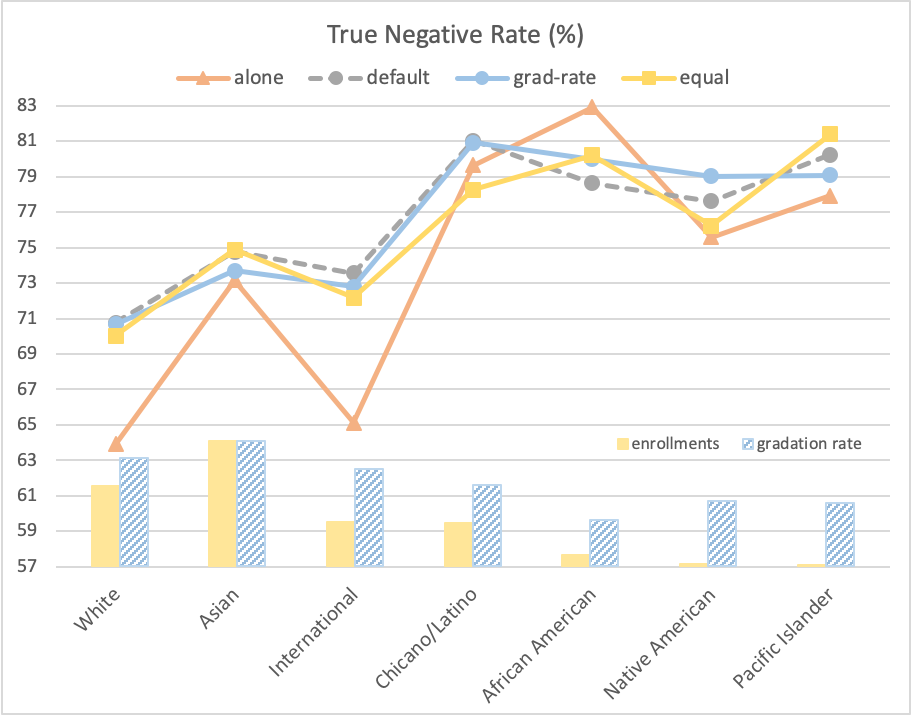}
		%\caption{Interpolation for Data 2}\label{Fig:Data2}
	\end{minipage}\hfill
	\begin{minipage}{0.33\textwidth}
		\centering
		\includegraphics[width=1\linewidth]{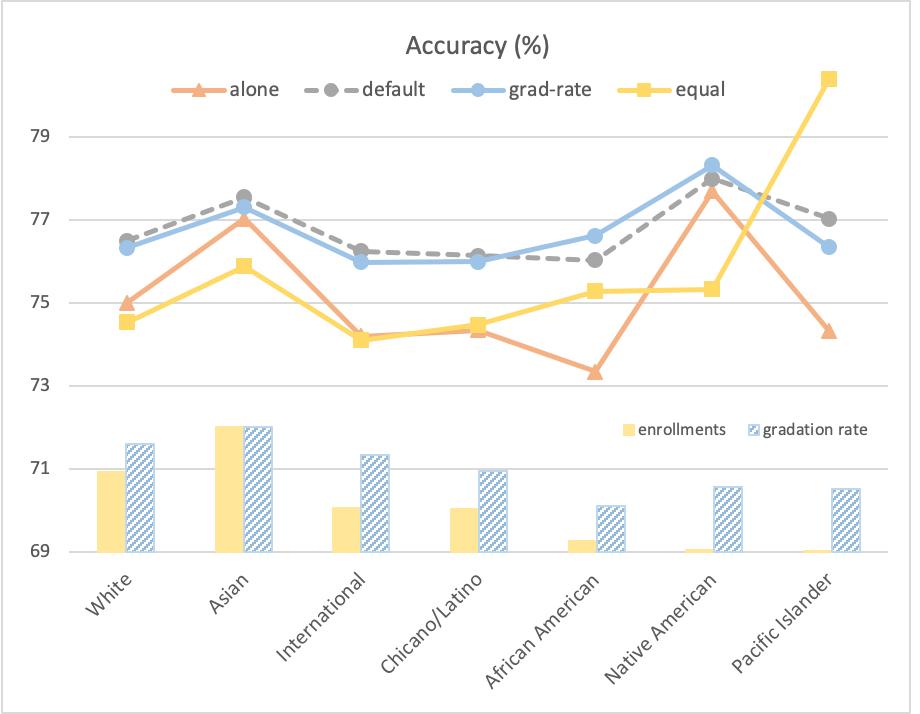}
		%\caption{Interpolation for Data 2}\label{Fig:Data2}
	\end{minipage}
	\caption{Evaluation results comparison between models with weighted loss by race}
	\label{wgh_race}
\end{figure*}
This equity of \textit{outcome} strategy is defined by weighing groups in reverse proportion to a longer-term educational outcome, such as the average graduation rate of all race groups $\bm{d}$. We defined the weighting function  $\lambda$ in equation (\ref{wgh_sample_eq}) as $\lambda(\bm{d})=1-\bm{d}$ this time because this formula tends to assign larger weights to races with lower graduation rates than using the inverse for our data, where $\bm{d}$ is the 6-year graduation rate vector of all race groups according to a diversity report\footnote{\label{note1}\url{https://diversity.berkeley.edu/reports-data/diversity-data-dashboard}} from the University. This strategy enables the model to focus on the groups for which an administration may want to focus attention on. Note that attention is proportional to a group's historic educational outcomes and not to its relative representation (i.e., not strictly related to being in the minority or majority, as with the instance balance condition). This equity of outcome condition is denoted by "grad-rate".

%Intuitively, if we intend to let the model pay most attention to a certain minority race group, we can set a very large weight for that race group and force the weights for the other groups to be very small, the model will tend to just learn from data of that race group. 
If a particular group is predicted to perform better when they are more represented, then would it follow that a group would perform the best if it was the only group contained in the training data? As a last condition, we define a strategy whereby training the grade prediction model is conducted separately on each group. This experiment setting is denoted by "alone".

Evaluation results on the three metrics are shown in Figure \ref{wgh_race}, with enrollments and graduation rate distributions across race in the bottom for reference\footnote{The height of each bar in the histogram represents the proportion of the corresponding value of a group to the maximum value of all the groups}. We found that separating race groups and training on them separately is not an ideal strategy for any group, as the accuracy decreased for all groups compared with training on the whole data, especially for Chicano/Latino, African American, and Pacific Islanders. Compared with results by the other strategies, TNR for most race groups was also the lowest under separated training, though TPR were slightly better than other strategies for the first three groups and Native American.
The salient discrepancies of results among race groups underscore that different patterns exist in the data of different race groups leading to disparities in the model's learning power and predictive power for each group.
%It could also be inferred that although we intended to let the model cater to certain groups based on their representativeness and graduation rate, which were designed towards equity of opportunity and equity of outcome, respectively, the samples outside the target group are also useful to the model's predictive power for the target group. 

Compared with "fairness through unawareness" (default), weighting samples to cater to race groups inversely proportional to graduation rates (grad-rate) helped to increase the TNR and accuracy for African American and Native American students, who historically have most struggled with on-time graduation\textsuperscript{\ref{note1}}, while almost maintained the group's TPR. This means more underserved students can be recommended appropriate remediation with minimal "underplacing" of others (i.e., FNR).%, which might contribute to future improvement in their graduation rates with proper intervention. 

\begin{figure*}[!htb]
	\begin{minipage}{0.33\textwidth}
		\centering
		\includegraphics[width=1\linewidth]{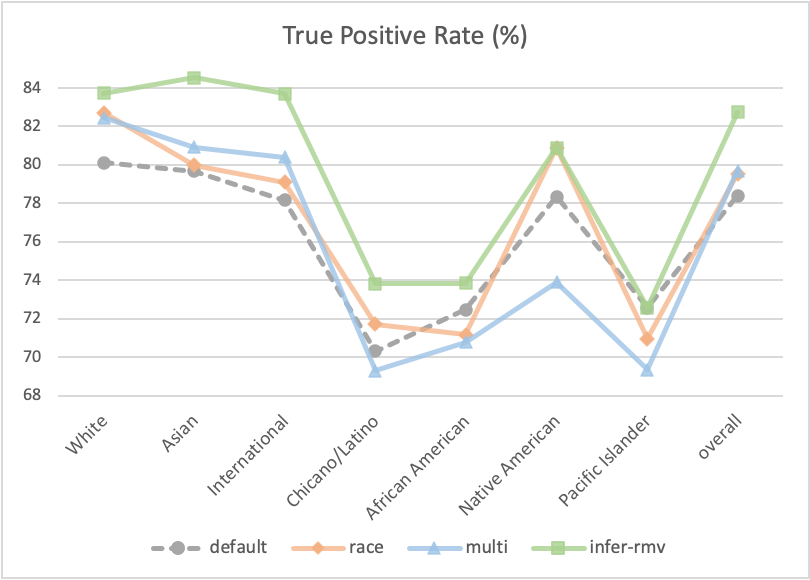}
		%\caption{Interpolation for Data 1}\label{Fig:Data1}
	\end{minipage}\hfill
	\begin{minipage}{0.33\textwidth}
		\centering
		\includegraphics[width=1\linewidth]{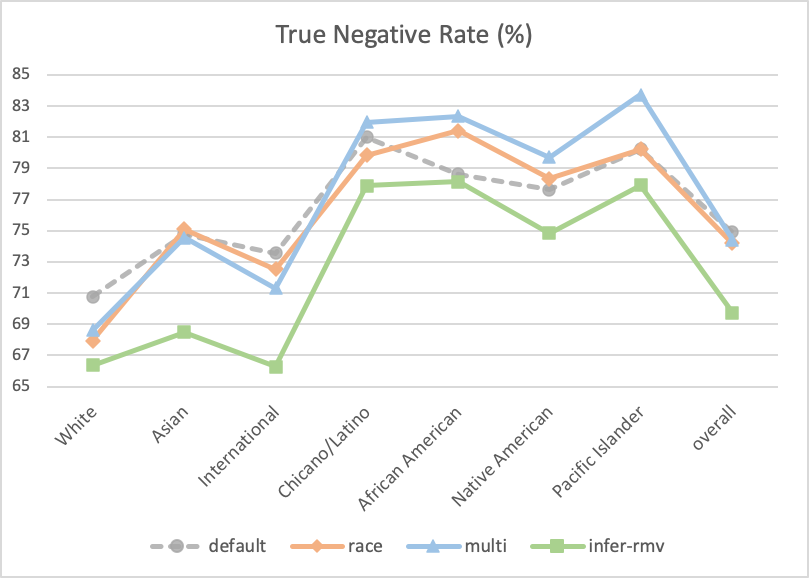}
		%\caption{Interpolation for Data 2}\label{Fig:Data2}
	\end{minipage}\hfill
	\begin{minipage}{0.33\textwidth}
		\centering
		\includegraphics[width=1\linewidth]{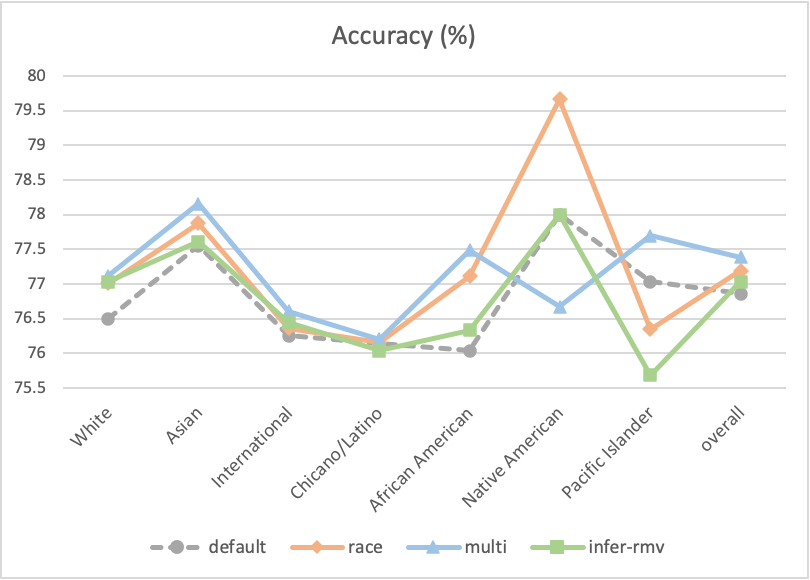}
		%\caption{Interpolation for Data 2}\label{Fig:Data2}
	\end{minipage}
	\caption{Results of models adding sensitive student attributes to the input}
	\label{feature_model}
\end{figure*} 

Balancing samples by race lowered the accuracy, TPR, and TNR for almost all race groups except Pacific Islanders, likely due to the number of Pacific Islanders only occupying around $0.2\%$ of the student population, far lower than other race groups. In light of the worse results by training only with Pacific Islanders, we can infer that the data pattern of this group is hard for the model to fit. 
Nevertheless, it is worth mentioning that weighting samples based on race population achieved the highest TPR, TNR, and accuracy for Pacific Islanders, a noticeable improvement compared with other strategies. This improvement is generally hard to attain due to the intrinsic tension between TPR and TNR. Also worthy of note is that, in terms of accuracy, both the default and grad-rate strategies perform substantially better than the alone strategy. This indicates that additional training instances provide a net positive impact on the strength of the grade prediction model over a smaller training set that is homogeneous with respect to race.

\subsection{The Impact of Sensitive Attributes}
\label{impact_features}

To evaluate the impact of sensitive attributes on the predictive power and fairness of the grade prediction model, we first added only student race information to the model input by concatenating a one-hot race representation to the grades input, followed by additional concatenated attributes, including gender, family income when admitted, entry status, and major(s), as we described in section "Data Construction Strategies". In addition, we evaluated the inference strategy proposed in the "Inference" section by deleting sensitive attributes from the model input in the prediction (inference) stage, hypothesizing that the learned feature embeddings might take away some sensitive attribute-related information from the learned course grade embeddings, thus potentially debiasing the course grade embedding.

Evaluation results (Figure \ref{feature_model}) reveal that adding sensitive attributes to the model input helped to increase the prediction accuracy for most race groups in general, which resonates with previous research find that the inclusion of race as a feature could improve overall accuracy in predicting college success \cite{kleinberg2018algorithmic}. However, as we fed more features to the model, the model became more discriminatory when it came to TPR and TNR. In particular, models incorporating race tended to increase the TPR for White, Asian, International, Chicano/Latino, and Native American students, while decreasing the TPR for African American and Pacific Islanders. The trend to discriminate against students from underrepresented groups became more obvious with respect to TPR when all the sensitive attributes were added to the model input. Such discrimination will lead to more underestimation for underrepresented groups (i.e., lower true positive rate and higher false negative rate). On the other hand, TNR for Chicano/Latino, African American, Native American, and Pacific Islanders increased when more sensitive attributes were included. The accompanied inverse trends of TPR and TNR as more attributes were included in the model input demonstrated the tension and tradeoff between TPR and TNR, which suggests it is hard for the model to improve detection of underperforming students without also "underplacing" other students. Our results also echo previous research showing that being aware of sensitive attributes might induce identity-based biases
in predictive analytics \cite{Barocas2019}. The large gap of TPR and TNR between majority groups and underrepresented groups is also observed in \citet{yu2020towards}. 

The post-preprocessing strategy of removing sensitive attributes from the model input in the prediction (inference) stage exhibited more extreme patterns of TPR and TNR for majority groups and underrepresented groups, where even all race groups received the highest TPR and the lowest TNR. Though counter-intuitive, the results suggest that the adjusted inference model tended to make overestimation on all groups of students at the expense of accurately predicting underperforming students. 

\subsection{Summary of Group Fairness Results}

Group fairness is defined by equalized odds \cite{hardt2016equality}. In our case, this would mean each student race group would have the same true positive and false positive rates. We evaluated the group fairness of the proposed strategies based on their TPR, TNR, and accuracy. The range (i.e., max value - min value) and standard deviation of each metric over all the groups could be deemed as a group fairness measure, lower values corresponding to less disparity between race groups and therefore greater fairness. Ideally, the range and standard deviation should be both 0 if group fairness is fully attained. Table \ref{fairness_table} presents the evaluation results of the proposed strategies on all race groups. We selected five models based on the proposed strategies, where "default" is the same original LSTM grade prediction model as seen in previous sections, "grad-rate(wgh)" and "equal(wgh)" are two loss weighting strategies by graduation rate and by population, respectively, which were discussed in section "Debias the Imbalanced Race Groups", "race(feature)" denotes the strategy of race being explicitly included in the model input discussed in section "The Impact of Protect Features", and "adversarial" refers to the adversarial learning strategy proposed in section "Model Training Strategy with Adversarial Learning". Note that: (1) All these strategies were also complemented with the "weighting loss by grade label" strategy for the sake of improvements to fairness and accuracy, as described in section "Debias the Imbalanced Grade Labels" and (2) The "alone" strategy from the section "Debias the Imbalanced Race Groups" and the "multi" and "infer-rmv" strategies from section "The Impact of Sensitive Features" are not included here due to poor performance in group fairness and overall accuracy shown in those sections' analyses.

% Table generated by Excel2LaTeX from sheet 'race_summary'

\begin{table*}[htbp]
	\small
	\caption{Performance of the four fairness and equity-based strategies compared to no strategy (default). Results are reported using the metrics of TPR, TNR, and accuracy for each race group with group fairness measures of range and standard deviation.}
	\begin{tabular}{cc|r|r|r|r|r|r|r|r|r|r}
		&       & %\multicolumn{1}{c|}{\rot{White}}
		\rot{White}
		& \rot{Asian} & \rot{International} & \rot{Chicano/Latino} & \rot{African American} & \rot{Native American} & \rot{Pacific Islander} & 
		\rot{Overall}&
		\rot{\textbf{Range}} & \rot{\textbf{STD}} \\
		\toprule
		\multirow{5}[0]{*}{TPR(\%)} & default & 80.10  & 79.67 & 78.16 & 70.31 & \textbf{72.46} & 78.34 & 72.58 & 78.39& 9.79  & 4.02 \\
		& grad-rate(wgh) & 79.89 & \textbf{80.07} & 78.27 & 70.09 & 71.96 & 77.71 & 72.58 &\textbf{79.82} & 9.98  & 4.13 \\
		& equal(wgh) & 77.36 & 76.65 & 75.49 & 69.93 & 68.51 & 74.52 & \textbf{79.03} & 79.46
		& 10.52 & 3.90 \\
		& race(feature) & \textbf{82.70}  & 79.99 & \textbf{79.10}  & \textbf{71.72} & 71.17 & \textbf{80.89} & 70.97 & 79.53 
		&11.73 & 5.14 \\
		& adversarial & 80.27 & 79.37 & 77.91 & 70.79 & 72.26 & 77.07 & 72.58 & 78.42
		&  \textbf{9.48} & \textbf{3.80} \\
		\hline
		\multirow{5}[0]{*}{TNR(\%)} & default & 70.76 & 74.76 & \textbf{73.56} & \textbf{81.01} & 78.63 & 77.62 & 80.23 & \textbf{74.91}
		& 10.25 & 3.75 \\
		& grad-rate(wgh) & 70.67 & 73.68 & 72.79 & 80.92 & 79.99 & \textbf{79.02} & 79.07 & 73.89
		& 10.25 & 4.09 \\
		& equal(wgh) & 70.04 & 74.89 & 72.17 & 78.27 & 80.20  & 76.22 & \textbf{81.40} & 73.69
		& 11.36 & 4.15 \\
		& race(feature) & 67.95 & \textbf{75.09} & 72.53 & 79.84 & \textbf{81.42} & 78.32 & 80.23 & 74.21
		& 13.47 & 4.89 \\
		& adversarial & \textbf{71.27} & 74.61 & 72.99 & 80.03 & 79.34 & 77.62 & 79.07 &74.75
		& \textbf{8.76} & \textbf{3.45} \\
		\hline
		\multirow{5}[0]{*}{Accuracy(\%)} & default & 76.50  & 77.55 & 76.25 & 76.14 & 76.04 & 78.00    & 77.03 &76.86
		& 1.96  & 0.76 \\
		& grad-rate(wgh) & 76.33 & 77.31 & 75.99 & 76.00    & 76.62 & 78.33 & 76.35 & 76.82 &2.34  & 0.85 \\
		& equal(wgh) & 74.54 & 75.89 & 74.11 & 74.48 & 75.29 & 75.33 & \textbf{80.41} &76.93
		& 6.30   & 2.16 \\
		& race(feature) & \textbf{77.01} & \textbf{77.88} & \textbf{76.36} & \textbf{76.15} & \textbf{77.11} & \textbf{79.67} & 76.35& \textbf{77.19}
		& 3.52  & 1.23\\
		& adversarial & 76.80  & 77.31 & 75.86 & 75.83 & 76.37 & 77.33 & 76.35 &76.81 & \textbf{1.50} & \textbf{0.62} \\
	\end{tabular}%
	\label{fairness_table}%
\end{table*}%

The adversarial learning strategy achieved all the minimums of range and standard deviation for TPR, TNR, and accuracy, demonstrating the best group fairness among all the compared strategies. Because the adversarial loss of predicting race is designed to ensure that the learned course grade embeddings and the hidden states of the model be minimally discriminative in terms of race, the model could learn bias-free deep representations from biased data. Though not the best strategy in terms of TPR, TNR, and accuracy, adversarial learning did not sacrifice much with respect to these metrics. No single strategy was always best with respect to those metrics; however, the strategy that most frequently scored the highest was the one in which race was included as a feature in the input. It was also most frequently the worst strategy with respect to measures of group fairness and always worse than the default in that regard, underscoring the inescapable but necessary trade-offs at play when designing for fairness \cite{kleinberg2016inherent, fazelpour2020algorithmic}.

\begin{figure}[h]
	
	\centering
	\includegraphics[width=\linewidth]{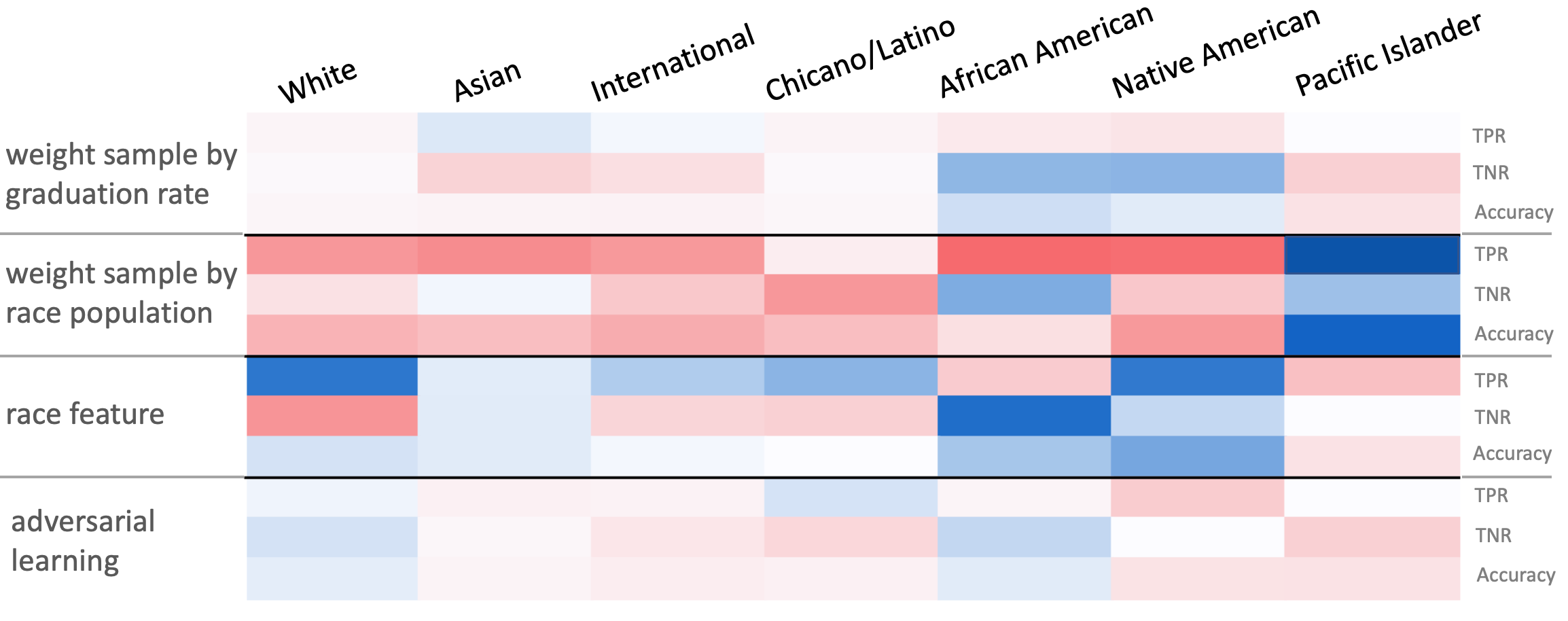}
	%\caption{Interpolation for Data 2}\label{Fig:Data2}
	\caption{Heat map of performance of the four fairness and equity-based strategies. A white background means performance was the same as the default (no-strategy), blue means performance was higher, and red means it was lower. Higher opacity represents higher magnitude.}
	\label{heatmap}
\end{figure}

We group predictive performance metrics together by strategy to more clearly observe how strongly different groups favor different strategies. Figure \ref{heatmap} depicts a heat map of the increase (blue) or decrease (red) in each metric relative to the default LSTM grade prediction model. Three of the four underrepresented groups were highly benefited by one of the four strategies; the Native American group was boosted in all three metrics by the "race feature" strategy. For the Pacific Islanders group,  balancing sample representation by race achieved the highest scores for the group in all metrics, handling the problem of the small population very well. Three of the four strategies helped to improve the TNR and accuracy for African American students. The debiased course grade representations learned by the adversarial learning strategy increased the TNR and accuracy for that group without much sacrifice of TPR. The comparatively lower TPR of African American students signifies that African American students tended to be underestimated.%, likely due to the biased course grade representations learned by the LSTM. %These results substantiate the power of the proposed strategies in reducing bias for underrepresented groups. %Specifically, more students who tend to struggle with their academic learning will be able to be detected accurately without "underplacing" potential well-performing students in those races. On the other hand, the overestimation problem on majority groups could also be solved for White and Asian students by adversarial learning and adding race feature to the model input, respectively. 

\section{Conclusions}
Fairness through unawareness was not most effective in achieving group fairness, as expected. However, presenting race explicitly to the input of the model led to the most unfair results out of all strategies. Instead, adversarial learning achieved the best fairness scores on all three metrics of TPR, TNR, and Accuracy.

Our equity of outcome approach, which sampled instances by group with inverse proportion to a historic educational outcome (e.g., graduation rate), was effective in boosting the predictive accuracy of most of the historically underserved groups. %It was also the strategy that achieved the highest overall accuracy, even besting the default (no-strategy) approach. 
Oversampling underrepresented groups helped in the case of Pacific Islanders, but was not effective for other groups and training exclusively on a group generally led to lower predictive performance for that group as compared to training on all groups. %Should students belonging to a different racial group receive different underlying models for course recommendation or other institutional support services? While this may be a controversial measure, the results highlight above in the summary section indicate that different models are most performant for different groups. Therefore, allowing students to opt into or out of a racially sensitive recommendation model may be worthy of consideration. 

We found grade label balancing to be an effective strategy for improving grade prediction TNR and TPR among underrepresented groups. This finding underscores the simple but important observation that a student group that mostly produces a minority label (e.g., lower grade) will likely be more poorly predicted than a group mostly producing the majority label. In educational contexts, where the majority grade is often higher than the minority grade, this will lead to perpetuating inequity, where students scoring lower will be worst served by the algorithms intended to help them. Grade label balancing mitigates this effect and further work is needed to develop additional best practices to address equity and fairness in the myriad of educational scenarios in which machine learning could otherwise widen achievement gaps.
%other applications: The use of algorithmic (data-driven and learning-based) decision making systems in domains ranging from judiciary (recidivism risk estimation) and banking (credit ratings and loan approval risk) to welfare (benefits eligibility) and insurance (accident risks) has raised numerous concerns about their fairness.

%\bibliography{sample-base}

\section*{Acknowledgements}
We thank the UC Berkeley Office of the Registrar, Office of Undergraduate Admissions, Office of Equity \& Inclusion, and Enterprise Data and Analytics for their anonymized enrollment and demographic data provisioning. The activities of this research study were approved by the Committee for the Protection of Human Subjects (protocol number: 2018-12-11671).
\bibliographystyle{ACM-Reference-Format}
\bibliography{sample-base}


\end{document}